**Design Principles for Self-forming Interfaces Enabling Stable Lithium Metal Anodes**

Yingying Zhu,[⌞,⌜] Vikram Pande,[⊥,⌜] Linsen Li,[⌞, †, *] Sam Pan,[†] Bohua Wen,[†] David Wang,[†] Venkatasubramanian Viswanathan,[⊥, *] Yet-Ming Chiang[†, *]

[⌞] Department of Chemical Engineering, Shanghai Electrochemical Energy Devices Research Center (SEED), Shanghai Jiao Tong University, Shanghai, China 200240

[†] Department of Materials Science & Engineering, Massachusetts Institute of Technology, Cambridge, MA 02139, USA

[⊥] Department of Mechanical Engineering, Carnegie Mellon University, Pittsburgh PA 15213，USA

[⌜] These authors contribute equally to the work.

* Corresponding authors

**Abstract**

The path toward Li-ion batteries with higher energy-densities will likely involve use of thin lithium metal (Li) anode (<50 μm thickness), whose cyclability today remains limited by dendrite formation and low Coulombic efficiency (CE). Previous studies have shown that the solid-electrolyte-interface (SEI) of the Li metal plays a crucial role in Li electrodeposition and stripping behavior. However, design rules for optimal SEIs on Li metal are not well established. Here, using integrated experimental and modeling studies on a series of structurally-similar SEI-modifying model compounds, we reveal the relationship between SEI compositions, Li deposition morphology and CE, and identify two key descriptors (ionicity and compactness) for high-performance SEIs. Using this understanding, we design a highly ionic and compact SEI that shows excellent cycling performance in high specific energy $LiCoO_2$-Li full cells at practical current densities. Our results provide guidance for rational design of the SEI to further improve Li metal anodes.

**Main Text**

Matching high-voltage oxide cathodes (> 4 V vs $Li^+$/Li) with thin lithium (Li) metal (<50 μm in thickness) anodes promises Li-ion batteries with specific energy exceeding 300 Wh $kg^{-1}$.[1] However, the cycle life of thin Li metal anodes is severely limited by short-circuits (i.e. "sudden death") caused by Li dendrite formation and low Coulombic efficiency (CE) as a result of side reactions between Li metal and electrolyte (i.e. "gradual death").[2-4] Recently published work has shown that dendrite formation may be suppressed to some extent by employing three-dimensional current collectors,[5,6] functionalized separators,[7-9] electrolyte additives,[10-13] surface coatings,[14-17], concentrated electrolytes,[18-20] and solid electrolytes.[21-23] With such improvements, it is urgent to address the low CE problem, for otherwise "gradual death" (running out of available Li or electrolyte dry-out) would likely occur before "sudden death" (short-circuit), and limit the cycle-life.

Li metal is highly reductive and reacts instantaneously with electrolyte constituents upon contact to form a surface film generally referred to as the solid-electrolyte-interface (SEI).[24,25] SEI formation consumes active $Li^+$ ions and leads to coulombic inefficiency. To minimize such loss, it is necessary for the SEI formation reaction to be self-limiting. It has also been shown recently that the microstructure and properties of SEI can impact the crystal growth behavior of Li metal during electro-deposition (charging of the cell)[26] and how the Li deposits are stripped during battery discharge.[27] Therefore, SEI tuning may be a promising strategy for improving Li metal anode performance. A variety of compounds, such as $Li_3PO_4$,[15] $LiF$,[28,29] $LiBr$,[30] $LiI$,[31] $LiNO_3$,[32] $Li_2S_8$,[33] $AlI_3$,[34] $SnI_2$,[35] $Al_2O_3$,[36] and $Cu_3N$,[37] have been used to modify the composition and morphology of the SEI and have been shown to be effective in improving the cycling performance of Li metal anode. However, these studies have typically been conducted with Li metal anodes of larger thickness (usually >250 μm). For high energy density, feasibility must be demonstrated with thin Li metal anode. It also appears that SEI tuning often follows a trial-and-error approach that results in incremental improvement. Thus, it is necessary to establish clear selection criteria for effective SEI modifiers.

Here, we first quantify the impact of Li metal thickness and CE on energy density and cycle-life of Li-metal rechargeable batteries. Then, using a model series of structurally-similar SEI-modifying compounds, we show the interrelationship between SEI compositions, Li deposition behavior, and CE. We identify two key descriptors (i.e. ionicity and compactness) for high performance SEIs using integrated experimental and modeling studies. Using this approach, electrolytes that result in a highly ionic and compact SEI enriched with LiF, $Li_2CO_3$, and $Li_2SO_3$, have been discovered, which form a dense Li film during electrodeposition (charge) and achieve both dendrite-free and high CE cycling. Li metal full batteries based on thin Li metal anodes (50 μm in thickness) and $LiCoO_2$ cathodes (theoretical areal capacity ~4.2 mAh cm$^{-2}$) demonstrate stable cycling exceeding 240 cycles (to 80% capacity retention) at practical C-rates (0.2 C charge/0.5 C discharge, 1 C ≈ 3.7 mA cm$^{-2}$). For even thinner Li metal anodes (20 μm thickness), full cells still cycle for 130 cycles. Our results provide guidance for rational selection and optimization of SEI modifiers to enable practical Li metal rechargeable batteries.

**Impact of electrode thickness and coulombic efficiency on energy-density, specific energy and cycle-life of Li metal batteries**

The calculated energy density, specific energy, and cycle life of Li metal batteries consisting of a high-area-capacity $LiCoO_2$ cathode (>4.2 mAh cm$^{-2}$) and a lithium metal anode of various thicknesses (i.e. 20 μm-thick Li corresponds to ~4.12 mAh cm$^{-2}$) are shown in **Figure 1**. The fraction of Li passed per cycle ($F_p$) can be calculated from the areal capacity of the cathode and the anode

$$F_p = \frac{Q_{cathode}}{Q_{cathode}+Q_{Li}} \qquad [1]$$

Here, $Q_{cathode}$ and $Q_{Li}$ are the area capacities of the cathode and Li metal anode, respectively; $F_p$ varies inversely with the percentage of Li excess of the battery, and naturally the battery reaches its highest energy density in the anode-free case, $F_p = 1$. Thus in **Figure 1** the energy density/specific energy decreases as $Fp$ decreases (Li excess increases), with the being calculated based on the mass and volume of the cathode, anode, current collectors, separator, and liquid electrolyte, and including a packaging factor, as given in detail in the **Supplementary Information**. The cycle life of the battery, n, may be predicted based on the $CE_{avg}$, which is the Coulombic efficiency averaged over the number of cycles until the all the available Li, $Q_{cathode} + Q_{Li}$, runs out. $CE_{avg}$ is calculated as follows:

$$CE_{avg} = 1 - \frac{Q_{cathode}+Q_{Li}}{nQ_{Li\,passed\,per\,cycle}} = 1 - \frac{Q_{cathode}+Q_{Li}}{nQ_{cathode}} \qquad [2]$$

($Q_{cathode} + Q_{Li}$)/***n*** *is* the average Li loss per cycle. Here we assume charging to 100% state-of-charge, and that CE loss occurs only at the Li metal anode, not at the cathode. Therefore, $Q_{Li\,passed\,per\,cycle}$ is equivalent to $Q_{cathode}$ during cycling. From **Equation 1** and **2**, the relation between $F_p$ and cycle life (*n*) can be determined as:

$$F_p = \frac{1}{n(1-CE_{avg})} \qquad [3]$$

The $F_p$ versus *n* plots in **Figure 1** are constructed based on **Equation 3** at several selected values of $CE_{avg}$. As shown in **Figure 1**, if the $CE_{avg}$ is as low as 80%, the battery cannot survive more than 60 cycles even with a 250-μm-thick Li metal anode (low $F_p$, large Li-excess). With a $CE_{avg}$ of 99.9%, even an anode-free battery can last more than 1000 cycles, approaching the cycle life of existing Li-ion batteries using graphite anodes. The cycle life of Li metal battery can be further increased to more than 2000 cycles when a thin Li metal anode (e.g. 20 μm or 50 μm) is employed. Due to the low density of Li metal (0.534 g cm$^{-3}$), this only leads to a small reduction in gravimetric energy density but a large reduction in volumetric energy density. Thus the increase in the thickness of the lithium metal foil improves the cycle life, but at the cost of specific energy and energy density. In this work, we demonstrate cycling of lithium metal foils with 20 and 50 μm thickness.

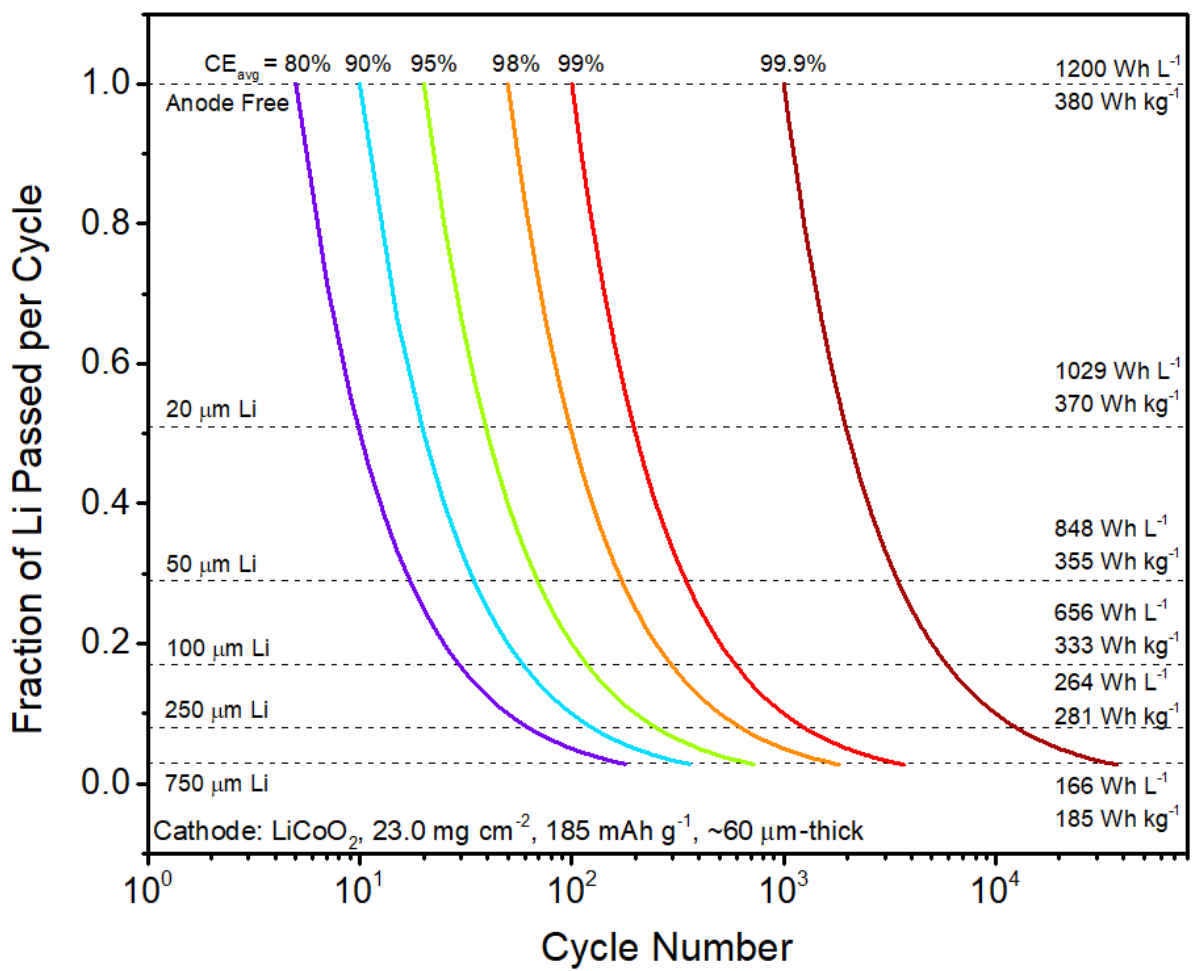

**Figure 1 | Prediction of energy density and cycle life of Li metal batteries.** The energy density, specific energy and cycle life are functions of the average Coulombic efficiency (CE$_{avg}$) and fraction of Li passed ($F_p$) per cycle. The mass and volume of the electrodes, current collectors, separator, electrolyte, and packaging are included in the calculation of gravimetric and volumetric energy density respectively, as detailed in **Supplementary Information**. The 20 μm, 50 μm, 100 μm, 250 and 750 μm-thick Li metal films are commercially available.

**Selecting SEI modifiers**

The central theme of this work is to establish the selection criteria for SEI modifiers that create a stable Li metal anode interface, and one which has the potential to self-heal upon formation of cracks. The spontaneous reaction between Li metal and iodine ($I_2$), which is historically used in primary batteries and leads to the formation of a Li$^+$ conducting layer and solid separator during operation,[38] offers a starting paradigm. However, due to the shuttle reactions associated with I$^-$/I$_3^-$/I$_2$, this limits the voltage capability of cathodes to less than 3.2 V.[39,40] Similar shuttle reactions are present for other halide species involving bromide and chloride at potentials below that of 4 V cathodes.[41] Among the halide series, this criterion leaves only LiF; the benefits of a LiF-rich SEI have been documented before.[12,29] However, direct addition of LiF as a salt in common organic electrolytes is not effective due to its extremely limited solubility (<0.002 mM in dimethyl carbonate). Therefore, we chose to explore an approach whereby LiF-rich SEIs are formed through intentional decomposition of fluorinated electrolyte constituents at the Li metal surface. Density-functional theory (DFT) calculations were used to probe reactions of a wide range of fluorinated organic compounds and lithium metal surface to determine their propensity to form a desired SEI (**Supplementary Fig. 1**). Note that these compounds can be considered either as solvents or additives depending on the amount added into the electrolyte. We begin by focusing on a series of structurally similar fluorinated organic compounds, namely fluoroethylene carbonate (FEC), di-fluoroethylene carbonate (DFEC) and 3,3,3-trifluoropropylenecarbonate (also known as trifluoromethyl ethylene carbonate, CF$_3$-EC) (see molecular structure in Figure 2a). The result from ethylene carbonate (EC) is also included for comparison. As shown in Figure 2b and 2c, FEC spontaneously decomposes to form LiF, unstable CO$^-$ anion and lithium salt of glycolaldehyde. On the other hand, DFEC decomposes partially upon ring opening, leading to formation of LiF and a large lithium alkoxide. Interestingly, CF$_3$EC, despite containing more F in its molecular structure, does not decompose to form LiF. These results clearly show that not every fluorinated organic solvent decomposes to form LiF at the Li metal surface. Interestingly, during the screening process for SEI modifiers, we have identified another approach to enriching LiF in the SEI, whereby the F atoms are extracted from the electrolyte salt LiPF$_6$. We discovered that 1,3,2-dioxathiolan-2,2-oxide (DTD, see molecular structure in Figure 2a) decomposes along with LiPF$_6$ to form LiF, PF$_5$, ethane-1,2-diolate (similar to a decomposition product of FEC) and SO$_2^{2-}$ anion (**Figure 2b & 2c**). These results indicate the potential for systematically tuning the inorganic and organic content in the SEI through spontaneous reactions between the organic electrolyte constituents and Li metal.

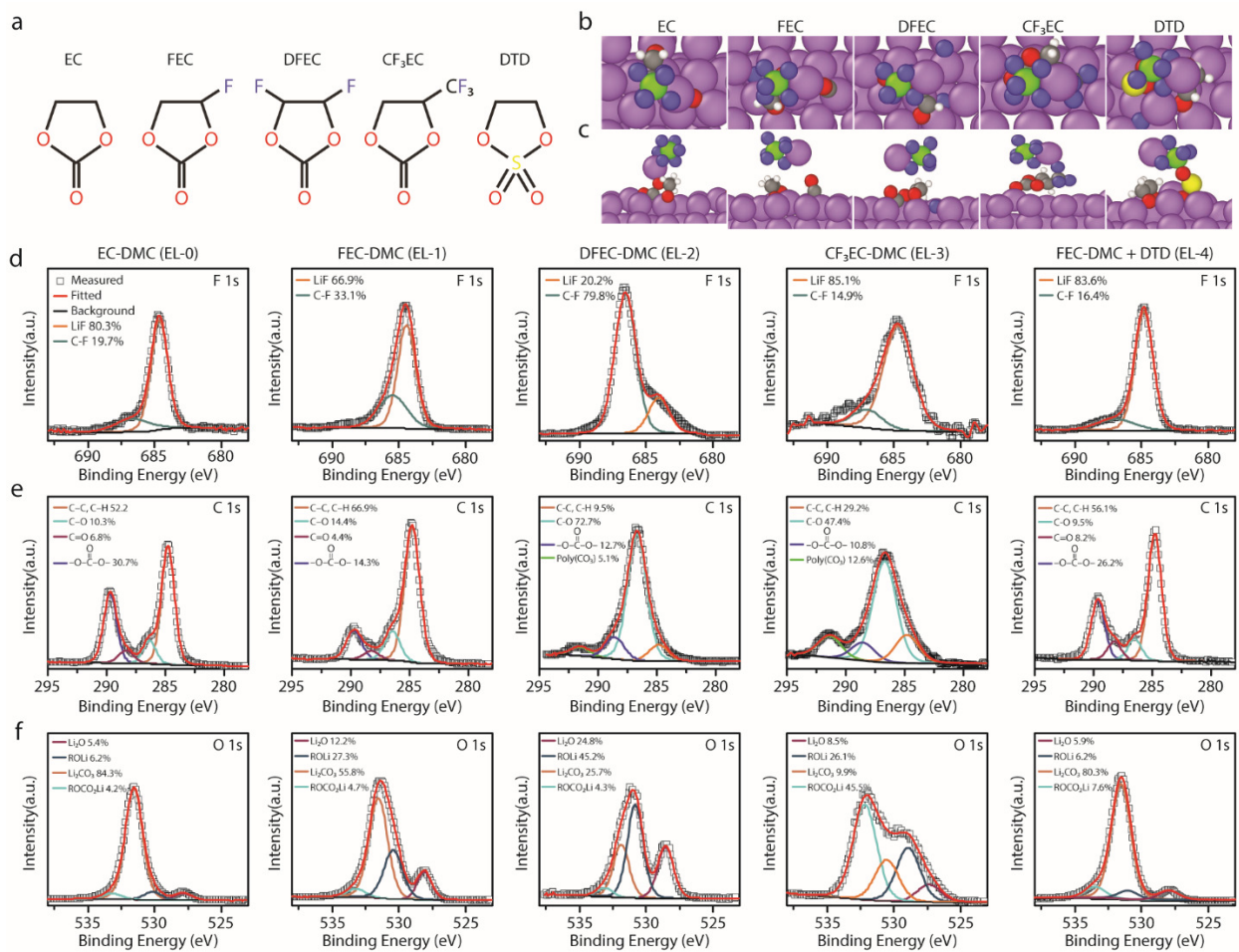

**Figure 2 | Decomposition of selected molecules at the Li surface studied by DFT calculations and XPS**. **a.** Molecular structure of ethylene carbonate (EC), fluoroethylene carbonate (FEC), di-fluoroethylene carbonate (DFEC), 3,3,3-trifluoropropylenecarbonate (CF$_3$EC), and 1,3,2-dioxathiolan-2,2-oxide (DTD). **b** and **c** are top and side views, respectively, of final decomposition products of EC, FEC, DFEC, CF$_3$EC, and DTD at Li (100) surface at the presence of LiPF$_6$, as predicted by DFT calculations. The purple atoms represent Li, red represent O, grey represent C, green represent P, blue represent F, yellow represent S and silver represent H. FEC and DTD break down completely, while DFEC decomposes partially and CF$_3$EC does not undergo significant breakdown. DTD also catalyzes the decomposition of LiPF$_6$ leading to formation of LiF. **d**, **e**, and **f** are narrow-scan XPS spectra of F 1s, C 1s, and O 1s, showing that DFEC and CF$_3$EC decomposes differently from FEC, despite their structural similarity.

To experimentally corroborate the DFT calculations, X-ray photoelectron spectroscopy (XPS) measurements were made to probe the SEIs of Li films deposited in electrolytes containing 1 M $LiPF_6$ dissolved in EC-DMC (EL-0, DMC = dimethyl carbonate), FEC-DMC (EL-1), DFEC-DMC (EL-2), $CF_3$EC-DMC (EL-3), and FEC-DMC + 3 wt% DTD (EL-4). Approximately the same amount of Li (~4.2 mAh cm$^{-2}$) was deposited on Li/Cu substrates (50 μm-thick Li, 15 μm-thick Cu) by charging $LiCoO_2$-Li cells to 4.5 V at 0.1 C. The deposited Li films were rinsed with fresh dimethyl carbonate (DMC) and dried under argon atmosphere before transferring to XPS measurements using an air-proof sample holder. Wide-scan XPS spectra of the SEIs (**Supplementary Fig. 2**) show that the F content in the SEI increases in the order of EL-0 (1.5 at.%), EL-3 (1.7 at.%), EL-1 (5.8 at.%), EL-4 (7.2 at.%), and EL-2 (10.6 at.%), which is in agreement with the trend predicted by the DFT calculations. Narrow-scan XPS spectra of F, C, and O are analyzed and the results are summarized in Figure 2d, 2e, and 2f, respectively. The F 1s spectra for all four SEIs show similar peaks that can be assigned to C-F ad LiF (**Figure 2d**). Interestingly, unlike the other three cases, there is more C-F than LiF in the SEI when EL-2 (DFEC-DMC) is used. Although XPS is a semi-quantitative method, it is safe to conclude that the SEI formed in EL-2 contains less LiF than that formed in EL-1 and EL-4. The F 1s spectrum of EL-2 may be explained by the partial decomposition of DFEC as suggested by the DFT calculation. The high intensities of the C-O peaks observed for the SEIs formed in EL-2 and EL-3, but not in EL-1 and EL-4 also support the conclusion that DFEC and $CF_3$EC do not decomposes as completely as FEC (Figure 2e). According to the O 1s spectra (Figure 2e), ROLi species are formed when FEC and DFEC are used, while a large amount of alkyl lithium carbonate ($ROCO_2Li$) is observed when $CF_3$EC is used, which confirms the DFT results. It is also observed in EL-4 that the presence of DTD in FEC-DMC promotes the formation of $Li_2CO_3$ over ROLi (compare EL-4 and EL-1).

For EL-4, the S 1s spectrum was also collected and analyzed **(Supplementary Fig. 3)**. Surprisingly, there is no S-containing species observed in the SEI. This may be explained by the decomposition rate of DTD versus FEC at Li surface. DFT calculations suggest that FEC decomposes directly on Li while DTD co-decomposes with $LiPF_6$. Thus we may expect the DTD decomposition reaction to be slower. Once the Li metal surface is passivated by the decomposition products of FEC (such as LiF), its reactivity toward DTD is significantly reduced. However, on the Li metal film that was cycled 20 times and 50 times, $ROSO_2Li$, $Li_2SO_4$, and $Li_2SO_3$ were indeed observed. This result suggests that S-containing species are gradually incorporated into the SEI during Li deposition/stripping cycles and may provide a healing function when the SEI cracks.

**SEI and Li metal electrodeposition behaviors**

To investigate the relationship between SEI composition and Li metal electrodeposition behavior, scanning electron microscopy (SEM) was performed on the Li film deposited in EL-0, EL-1, EL-3, and EL-4 electrolytes using the LiCoO$_2$-Li cells. The same Li films on which the XPS measurements were made were examined by SEM. The SEM used for this study was installed inside an argon-filled glovebox so that the samples were never exposed to air. Top-view SEM images showed that Li particles of several microns were deposited in EL-0, EL-1, and EL-4 (**Figure 3a**, **3b**, and **3d**). Smaller particles were observed in EL-3 (**Figure 3c**). The difference in Li electrodeposition behavior among the four electrolytes is most clearly seen in the cross-sectional SEM images in **Figure 3e-3h.** The deposited Li films were clearly thicker in the EL-0 (~35 µm) and EL-3 electrolytes (~37 µm) than EL-1 (~25 µm), and EL-4 electrolytes (~26 µm), despite having the same areal capacity or areal mass. Since the deposited capacity of 4.2 mAh cm$^{-2}$ corresponds to a thickness of ~20 µm if we assume that the Li film is fully dense, we can estimate the porosity of the deposited Li film in the four electrolytes using the thickness of the deposited Li observed in cross-sectional SEM. These densities are 43% for the EL-0, 20% for the EL-1, 54% for the EL-3, and 22% for the EL4. It appears that the LiF and Li$_2$CO$_3$-rich SEIs formed in the EL-1 and EL-4 electrolytes promote the deposition of dense Li films and with less of a whisker-like Li morphology. This morphology appears to correspond to a high CE. It has been shown that whisker-like Li particles are prone to cracking during stripping and thereby lose contact to become "dead Li",[27] which likely leads to a low CE. More whisker-like Li particles and a wavy surface of the deposited Li film were indeed observed when a thick Li foil was used as the counter electrode instead of the LiCoO$_2$ electrodes (**Supplementary Fig. 4**).

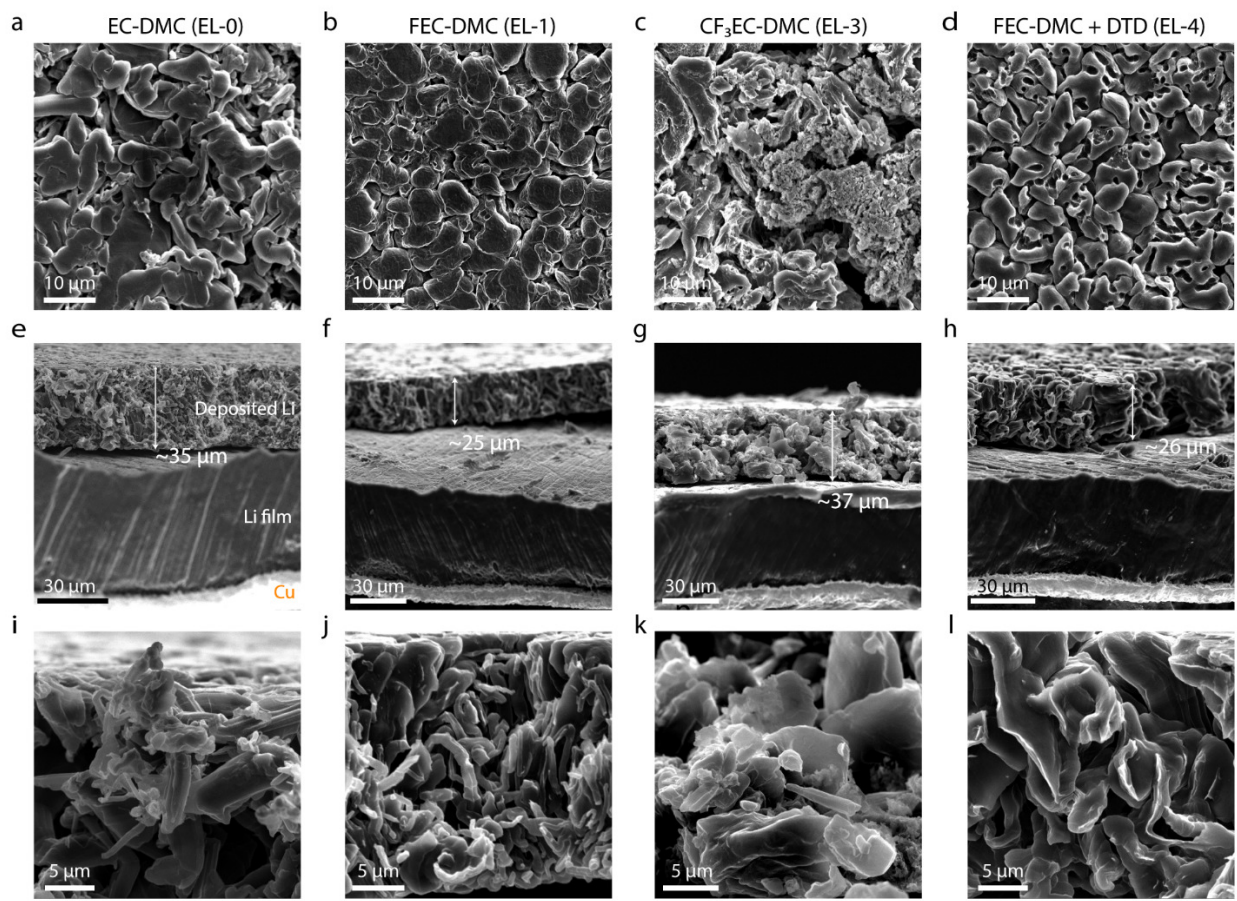

**Figure 3 | SEM characterization of the deposited Li film on the Li/Cu substrates. a**, **b**, **c**, and **d** are top-view SEM images of the deposited Li films on the Li/Cu substrates (50 μm-thick Li, 15 μm-thick Cu) in 1 M $LiPF_6$ EC-DMC (EL-0), 1 M $LiPF_6$ FEC-DMC (EL-1), 1 M $LiPF_6$ $CF_3$EC-DMC (EL-3), and 1 M $LiPF_6$ FEC-DMC + DTD (EL-4), respectively. The Li films were deposited in $LiCoO_2$-Li cells by charging at 0.1 C to 4.5 V vs $Li^+$/Li. The same amount of Li (~4.2 mAh $cm^{-2}$) was deposited for all five cases. Micro-sized Li particles were observed in EL-0, 1, and 4. Smaller particles were observed in EL-3. **e**, **f**, **g**, and **h** are the corresponding cross-sectional SEM images, showing three-layer structure consisting of the deposited Li, the original 50 μm-thick Li, and the underlying Cu substrate. The deposited Li films are thicker in EL-0 and EL-3 than EL-1, and EL-4. Panel **i**, **j**, **k**, and **l** are cross-sectional SEM images at higher magnifications.

**Coulombic efficiency measurements using asymmetric Li-Li cells**

In order to accurately measure Coulombic efficiency during plating and stripping, it is necessary to have a limited lithium source so that the loss of working lithium can be traced. This can be done with a full cell using an intercalation cathode, but any losses at the positive electrode may be difficult to separate from those occurring at the lithium metal electrode. We developed an asymmetric Li-Li cell test that is able to accurately quantify the average Coulombic efficiency occurring over a number of cycles, which we denote $CE_{avg}$ (**Figure 4a**). The asymmetric cell consists of a two Li metal electrodes, one of which has a low area capacity that is systematically consumed during cycling. In the present study we used a 20 μm-thick Li film coated on a copper foil ($Q_{Li}$ = 4.12 mAh cm$^{-2}$) as this working electrode, while the counter electrode is a 750 μm-thick Li foil with a large excess of capacity. The two electrodes were assembled into a coin-cell with a polyethylene separator and liquid electrolyte. In the first half-cycle, a known amount of Li, in this instance 3.0 mAh cm$^{-2}$ per cycle ($Q_{Li}$ passed per cycle), is deposited on the thin working electrode (here, at a current density of 0.6 mA cm$^{-2}$). The same 3.0 mAh cm$^{-2}$ is then stripped from the working electrode. With each successive cycle, the same $Q_{Li}$ passed per cycle is stripped and plated. Any Coulombic inefficiency erodes the initial 20 μm-thick Li film on the working electrode. Barring a short-circuit event, the original thin Li electrode is gradually consumed by the side reactions, either forming SEI or being isolated by SEI during cycling (**Figure 4b**) and forming so-called "dead Li." When all the initial Li at the working electrode is consumed, a voltage spike is observed. Three selected examples are shown in **Figure 4c**. Crucially, the appearance of voltage spike (denoted by the black arrows in **Figure 4c**) shows that short-circuits are absent. The average Coulombic efficiency over the number of cycles the cell experienced up to the voltage spike is $CE_{avg}$, and is calculated from an equation similar to **Equation 2**:

$$CE_{avg} = 1 - \left(\frac{Q_{total\ Li}}{n}\right) \cdot \left(\frac{1}{Q_{Li\ passed\ per\ cycle}}\right) = 1 - \frac{Q_{Li\ passed\ per\ cycle} + Q_{Li}}{nQ_{Li\ passed\ per\ cycle}} \qquad [4]$$

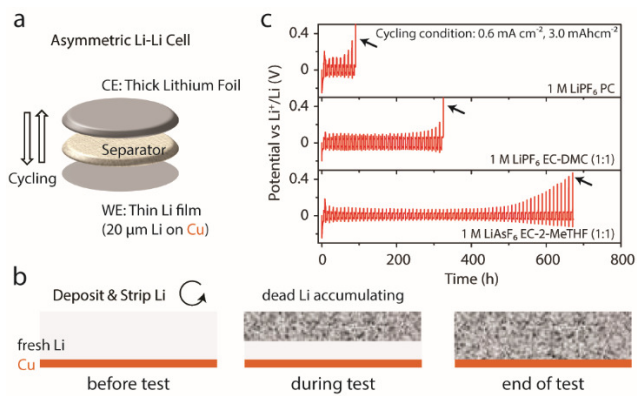

**Figure 4 | Asymmetric Li-Li cell design and test. a.** Schematic illustration of an asymmetric Li-Li cell consisting of a thin Li metal electrode (working electrode, WE), a separator, and a thick Li metal electrode (counter electrode, CE). In the first half-cycle, a fixed amount of Li is electrochemically deposited onto the thin Li electrode, and then this amount is stripped and deposited repeatedly. **b.** Evolution of the Li film on the thin Li electrode (WE) during the test. The Li originally coated on the Cu substrate was gradually consumed by the side reactions, and some becomes "dead Li" insulated by a thick SEI layer. **c.** Voltage curves of three selected examples of the Li-Li asymmetric cell tests. The cycling current density was 0.6 mA cm$^{-2}$. The cycling area-capacity is 3.0 mAh cm$^{-2}$. Li was first deposited on the thin Li electrode and then stripped. The final voltage spikes denoted by the black arrows indicate the end of the tests when there is no Li available for stripping anymore and the absence of short-circuits during the tests. Longer cycle time before the voltage spikes indicate higher CE$_{avg}$, based on Equation 4.

We tested the effectiveness of this approach by measuring the CE$_{avg}$ of the thin Li anodes in several selected electrolytes reported in previous literature. The results are summarized in **Supplementary Table 2.** Different values of CE$_{avg}$ were clearly observed for different electrolytes. The Li metal anodes were reported to cycle well in 1 M LiTFSI EC/tetrahydropyran,[42] 1 M LiAsF$_6$ EC/2-methyl-tetrahydropyran,[43] and 1 M LiTFSI 1,3-dioxolane/1,2-dimethoxyethane (1:1 v) + 1 wt% LiNO$_3$ electrolytes[44] but poorly in 1 M LiPF$_6$ propylene carbonate.[45] Correspondingly, high CE$_{avg}$ values were observed for the former three and a low CE$_{avg}$ was found for the latter. The CE$_{avg}$ of the thin Li metal anodes in the electrolytes EL-0 to 4 was then measured. The commonly used LiPF$_6$ EC-DMC electrolyte (EL-0) showed a CE$_{avg}$ of 92.6%. Replacing EC with a fluorinated EC such as FEC and DFEC significantly improved the CE$_{avg}$ to 97.0% and 96.2%, respectively. However, this beneficial effect was not observed for CF$_3$EC, which showed a very low CE$_{avg}$ of 20.9%. The EL-4 electrolyte with 3wt% DTD additive showed the highest CE$_{avg}$ of 97.9%. High $CE_{avg}$ was observed for the cases where the SEI contained more ionic compounds and the deposited Li was denser.

**Table 1 | Average Coulombic efficiency of the thin Li anodes in different electrolytes**

| Lithium Salt (1 M) | Solvents & Additives | $CE_{avg}$ |
|---|---|---|
| $LiPF_6$ | EC-DMC (1:1 v), EL-0 | 92.6% |
| $LiPF_6$ | FEC-DMC (1:1 v), EL-1 | 97.0% |
| $LiPF_6$ | DFEC-DMC (1:1 v), EL-2 | 96.2% |
| $LiPF_6$ | $CF_3$EC-DMC (1:1 v), EL-3 | 20.9% |
| $LiPF_6$ | FEC-DMC (1:1 v) + DTD, EL-4 | 97.9% |

**Test conditions**: 0.6 mA cm$^{-2}$ current density, 3.0 mAh cm$^{-2}$ cycling areal capacity.

The asymmetric Li-Li cell test provides a useful platform to evaluate and compare the effectiveness of different electrolyte components (salts, solvents, and additives). It is able to quantify both the CE and area-specific impedance (ASR) of the Li metal electrodes, whereas the conventional symmetric Li-Li cell tests can only quantify ASR. Furthermore, "soft" short circuits can be difficult to differentiate from a low ASR in symmetric Li-Li cells, whereas the asymmetric configuration yields unrealistically high CE (i.e. outliers) when short circuits are present. The asymmetric Li-Li cell test is also more directly relevant to practical applications than the widely used asymmetric Li-Cu (or Ni, or stainless steel) cell test in which Li is deposited on bare metal current collector, since the use of thin Li anodes in a full cell in most instances will provide a better compromise between energy density and cycle life than the so-called "anode-free" configuration where deposition occurs on metal current collector.

The cycling performance of the thin Li metal anodes in the presence of the different electrolytes and SEIs was further evaluated in Li metal full cells consisting of a "high-voltage" $LiCoO_2$ cathode (area-capacity ~4.2 mAh cm$^{-2}$ when charged to 4.5 V vs Li$^+$/Li), and a 20 μm-thick Li anode coated on a copper foil (areal capacity ~4.12 mAh cm$^{-2}$), and a single-layer polyethylene separator. 2025-type coin cell cases made of 316L grade stainless steel and a single-layer polyethylene separator, were used for the tests. The full cells were cycled at 0.2 C charge-0.5 C discharge between 4.5 and 3.0 V. Three formation cycles were performed at 0.1 C before the long-term cycling tests. The $LiCoO_2$-Li (20 μm) battery with the EL-0 electrolyte showed rapid capacity fade (**Figure 5a** and **5b**). The cycling performance was improved with the EL-1 electrolyte. The best cycling performance was achieved with the EL-4 electrolyte containing the DTD additive. 80% of the initial capacity was retained for 131 cycles (**Figure 5b**). It is also interesting to compare the CE of the three cells. The best-performing cell using the EL-4 electrolyte provides higher CE over those using the EL-0 and EL-1 electrolytes (**Figure 5c**). When a 50-μm Li metal anode was employed instead of the 20-μm Li metal anode, the $LiCoO_2$-Li full battery retained more than 80% of the initial capacity for 246 cycles (**Figure 5d**). This shows that thickness of lithium provides an effective way to strike an application dependent trade-off between cycle life and specific energy.

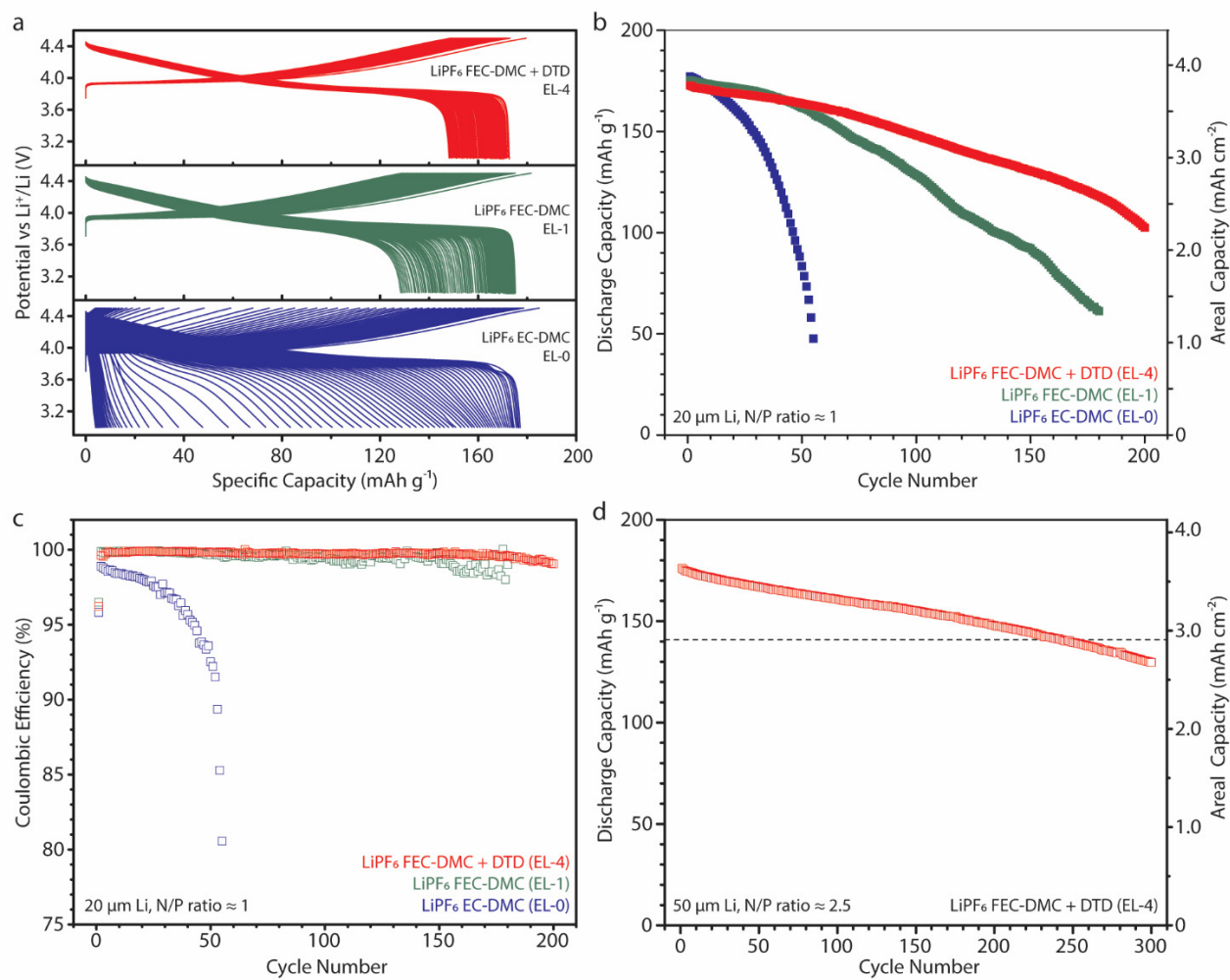

**Figure 5 | Electrochemical tests of $LiCoO_2$-Li full batteries. a.** Voltage profiles of three $LiCoO_2$-Li full batteries containing 1 M $LiPF_6$ EC-DMC (EL-0), 1 M $LiPF_6$ FEC-DMC (EL-1), and 1 M $LiPF_6$ FEC-DMC + DTD (EL-4). The batteries were first cycled at 0.1 C for three cycles and then charged at 0.2 C and discharged at 0.5 C repeatedly between 4.5–3.0 V. The $LiCoO_2$ electrodes are ~23 mg cm$^{-2}$ in mass

loading and ~4.2 mAh cm$^{-2}$ in areal capacity. The Li electrodes are ~20 μm in thickness, ~4.12 mAh cm$^{-2}$ in areal capacity. Data from the cycle 1 to 100 is shown. **b.** Cycling performance and **c.** Coulombic efficiency of the LiCoO$_2$-Li full batteries containing the three different electrolytes. EL-4 provides better cycling performance and higher Coulombic efficiency. **d.** Cycling performance of a LiCoO$_2$-Li full battery containing the 1 M LiPF$_6$ FEC-DMC + DTD (EL-4) electrolyte with a Li metal anode of 50 μm in thickness, ~10.30 mAh cm$^{-2}$ in areal capacity. 80% of the initial capacity was retained after 246 cycles.

It is important to note that the present results were obtained in the absence of any significant applied pressure. Pressure applied to lithium metal cells can dramatically improve their cycling stability. Unlike intercalation electrodes that undergo a comparatively small volume change during cycling, Li metal electrodes undergo a large volume change as they dissolve during discharge, which effectively reduces the stack pressure applied to the electrodes and likely changes the Li electrodeposition behavior. Observations made in the present external-pressure-free cells may differ from what occurs in wound cells that are under winding pressure, or stacked cells that may be under pressure from the cell casing or externally applied pressure. We expect that the present results would be further improved in cell configurations with significant applied pressure.

The performance of the Li metal full batteries with the EL-4 electrolyte is among the best reported for Li metal rechargeable batteries. Under the same test conditions, the EL-4 electrolyte provides even better cycling performance and CE than the previously reported dual salt electrolyte[10] (**Supplementary Fig. 6**). We further compared our results with others using an updated version of the "ARPA-E plot" previously shown in ref. 1 (**Figure 6**). The data points in the "ARPA-E plot" are analyzed in terms of four parameters: cumulative capacity plated (Ah cm$^{-2}$), plated current density (mA cm$^{-2}$), per-cycle areal capacity (mAh cm$^{-2}$), and fraction of Li passed per cycle ($F_p$). Compared with other previous work using liquid electrolytes (point 15 to 35), our work stands out for high $F_p$ ($F_p$ = 0.42 for point 36; $F_p$ = 0.23 for point 37) and relatively high per-cycle areal capacity (>3.2 mAh cm$^{-2}$), which means higher energy density based on the analysis shown in **Figure 1**. The data points in red color (i.e. low $F_p$ values) are from cells having relatively low overall energy density and far from meeting the DOE goals (point 1 and 2, green color, $F_p$ = 0.8). In this work, we have demonstrated cells with a specific energy of 370 Wh/kg lasting 130 cycles and cells with a specific energy of 355 Wh/kg lasting 246 cycles at practical current densities and cathode loadings. We also use 20 *μ*m thin lithium and have more than twice the lithium utilization (~42%) compared to any previous published work. Further, the voltage profiles clearly rule out the possibility of a soft short. Finally, our cells exhibit linear capacity fade, distinct from the typically observed non-linear drop in capacity after few cycles.

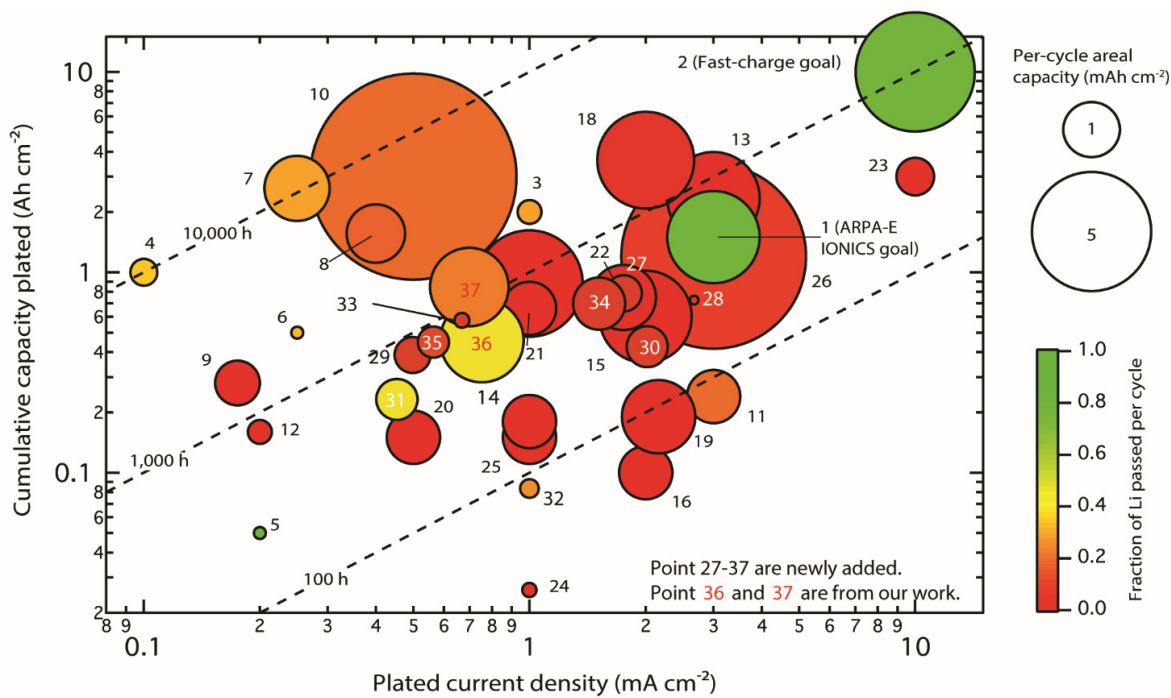

**Figure 6 | Comparison of cycling performance of Li metal anode.** This figure is an **updated version** of the one in Reference 1 ("Points 1 and 2 are DOE goals, 3–6 are for LiPON thin-film cells, 7–9 are PEO-based solid polymer electrolytes, 10–12 are solid inorganic separators, 13 and 14 are custom nanostructures, and 15–26 are liquid electrolytes"). Data points 27-37 are newly added based on recent literatures, which are from liquid electrolytes as well, just like points 15-26. See Supplementary Table 2 of this manuscript for references for each points. Data point 36 and 37 are from our work. Data point 36 and 37 are from $LiCoO_2$ (~4.2 mAh cm$^{-2}$)-Li (20 μm) and $LiCoO_2$ (~4.2 mAh cm$^{-2}$)-Li (50 μm) full cells, respectively. EL-4 electrolyte (1 M $LiPF_6$ FEC-DMC + 3 wt% DTD) were used. The cycling data were shown in Figure 5. Compared with other previous work using liquid electrolytes, our work stands out for higher fraction of Li passed per cycle ($F_p$ = 0.42 for point 36; $F_p$ = 0.23 for point 37), which means higher energy density based on the analysis shown in **Figure 1** of this manuscript.

**Descriptors for Effective SEIs**

In order to formulate the structure-property relationship between the electrolyte components and their function in full cells, we identify two descriptors that are key for a high performance, self-formed SEI.

The two key factors are (i) the ionicity of the SEI, which is needed for ensuring low electronic conductivity[46] and (ii) the porosity of the SEI, which will control the morphology and packing density of the inorganic and organic phases. Here, we propose to use the number of electrons transferred from the lithium surface to the electrolyte molecule as a descriptor for the iconicity of the SEI. This descriptor is intuitive as the more ionic the SEI is, the more electrons should be transferred from the lithium surface, as shown in **Table 2**. On the other hand, describing the porosity and morphology of the SEI is a more challenging problem. We propose that the volume of the organic species left behind from the decomposition is a good descriptor to describe the ability to form a compact SEI. This volume can be quantified by the Bader Volume of the largest SEI species, typically the organometallic salt component. More details on the rationale behind the two descriptors is discussed in the **Supplementary Information**. These two descriptors, used in conjunction, can rationalize the experimentally observed trends. Based on the descriptor for the ionicity of the SEI, among the present fluorinated solvents, the trend is $CF_3EC \ll$ FEC < DFEC. Using just this descriptor, we would conclude that DFEC leads to a more ionic SEI. However, when comparing the descriptor for compactness of the SEI, the incomplete decomposition of DFEC leads to much larger moieties for DFEC than FEC. These two factors taken together indicates that FEC should fare better due to a combination of an ionic and compact SEI.

The case of DTD is special and it suggests a new strategy to enrich the LiF content in the SEI. Unlike the other electrolyte molecules considered, DTD decomposes along with $LiPF_6$ to form LiF, $PF_5$, ethane-1,2-diolate (similar to FEC) and $SO_2^{2-}$ anion. The latter eventually leads to the formation of $Li_2SO_4$, $Li_2SO_3$ and $ROSO_2Li$ as observed in the experiments. For the two descriptors identified, DTD leads to the formation of a more ionic SEI while maintaining high density quite similar to FEC. We attribute the improved behavior seen in FEC-DTD case to increased LiF and $Li_2CO_3$ formation from the DTD addition as well [4]as the formation of $Li_2SO_3$ as confirmed by the XPS results. These results suggest that these two descriptors together can provide rational design principles for the selection of SEI forming compounds.

**Table 2 | Theoretical Descriptors for Solvent decomposition on Li metal**

| Solvent | Electrons Transferred | Bader Volume of Largest SEI Specie ($Å^3$) |
|---|---|---|
| DMC | 2.0 | 145 |
| EC | 2.5 | 132 |
| FEC | 3.3 | 89 |
| DFEC | 4.4 | 125 |
| CF$_3$EC | 0.5 | 169 |
| DTD | 4.1 | 97 |

**Conclusions**

We have shown that using thin Li metal anode and having high CE are two prerequisites for high energy-density and long cycle-life Li metal batteries. It is discovered that the decomposition of FEC and DTD (at the presence of LiPF$_6$) at Li metal surface enriched the SEI with ionic compounds such as LiF, Li$_2$CO$_3$, and Li$_2$SO$_3$ rather than other organic species, whereas structurally similar DFEC and CF$_3$-EC decompose and modify the SEI differently. A compact and ionic SEI promotes dense Li electrodeposition and high CE during cycling. LiCoO$_2$-Li full cells demonstrated stable cycling performance over 240 cycles at practically relevant areal-capacity (>3.2 mAh cm$^{-2}$) and C-rates (0.2 C charge/0.5 C discharge, 1 C = 3.7 mA cm$^{-2}$). This work establishes a methodology for rational selection and optimization of the SEI modifiers to further improve the cycling performance of Li metal anode.

**Methods**

**Materials.** The 20 μm-thick lithium film coated on Cu foil was purchased from Honjo, Japan. The 50 μm-thick lithium film coated on Cu foil was purchased from China Lithium Energy Co., Ltd. The electrolytes were prepared by (Shanghai Songjing New-Energy Technology) using battery-grade reagents except EL-2 because DFEC was only available at 95% purity (from BOCSCI Inc.). The LiCoO$_2$ electrode sheets were prepared by coating the slurry of LiCoO$_2$ cathode powder (LC-95, Hunan Shanshan), carbon black, and PVDF (weight ratio 96:2:2) on aluminum foils. The electrodes were calendared to ~60 μm. The mass loading of LiCoO$_2$ was ~23 mg cm$^{-2}$.

**Characterizations.** XPS measurements were carried out using a Kratos X-ray photoelectron spectrometer. To avoid electrode contaminations caused by exposure to air, the Li electrodes were rinsed using fresh dimethyl carbonate, dried, sealed in a specialized holder inside the Ar-filled glovebox, and then transferred into the chamber of the XPS instrument. SEM characterizations were carried out using a Phenom-Pro scanning electron microscope installed inside an Ar-filled glovebox. The cross-section of the deposited Li films were prepared by slowly tearing the electrodes. Using a razor blade to cut the electrodes is not advised because it easily deforms the soft Li metal and appears to "smooth out" the deposited Li films.

**Electrochemical measurements.** Electrochemical performances were measured using CR2025 coin-type cells. The Li metal full batteries were assembled using a LiCoO$_2$ cathode, a thin Li metal anode (20 or 50 μm in thickness) and one piece of polyethylene separator (Tonen), and <75 μL electrolyte per cell. Electrochemical tests were performed using battery cyclers (Shenzhen Neware, BTS4000-5V, 10/1.0 mA version, 0.05% current/voltage resolution). The coin cells were cycled inside temperature chambers set at 30 ± 0.2 °C. The lithium metal full batteries were first cycled at 0.1 C between 4.5–3.0 V for three cycles and then charged at 0.2 C and discharged at 0.5 C repeatedly. 1 C equals to ~3.7 mA cm$^{-2}$.

**DFT simulations.** Self-Consistent DFT calculations were performed using the real space projector-augmented wave method[47,48] implemented in the GPAW code[49,50] and employing the PBE exchange correlation functional[51]. We performed the DTF calculations on the Li(100), (110) and (111) surfaces. The Li surfaces comprised of four layers with the bottom two layers constrained at the bulk lattice constants. Each layer consisted of 3×3 Li unit cell. The solvent molecule along with Li$^+$ and PF$_6^-$ ions were placed on top of Li surface and the structure was allowed to relax to determine the decomposition. Periodic boundary conditions were used for *x* and *y* directions and a vacuum of 10 Å was used in the *z* direction perpendicular to the surface on both sides of the slab. A real-space grid spacing of 0.16 Å was used and the Brillouin zone was sampled using the Monkhorst Pack scheme[52] with a k-point grid of (6×6×1). All simulations were converged to a force < 0.05 eV Å$^{-1}$. Lastly Bader analysis[53] was used to determine the amount of charge transferred from the lithium to the solvent during the decomposition and also volumes of the various decomposed species.

**Data availability.** The supporting data for the included charts/graphs within this paper, as well as other findings from this study, are available from the corresponding author upon reasonable request.

**Acknowledgements.** This work was supported by the Assistant Secretary for Energy Efficiency and Renewable Energy, Office of Vehicle Technologies of the US Department of Energy (DOE) through the Advanced Battery Materials Research (BMR) Program under contract no. DE-EE0007810 and supported by the start-up fund of Shanghai Jiao Tong University (to L.S.L).

**Author contributions.** L.S.L., V.V., and Y.-M.C. conceived and supervised the research. Y.Y.Z. and L.S.L. performed the experiments with the help from M.S.P., B.H.W., D.W.. V.P. carried out the DFT simulations and analysis. L.S.L., V.P., V.V., and Y.-M.C prepared the manuscript with input from other co-authors.

**Reference**

1. Albertus, P., Babinec, S., Litzelman, S. & Newman, A. Status and challenges in enabling the lithium metal electrode for high-energy and low-cost rechargeable batteries. *Nature Energy* **3**, 16-21 (2018).
2. Zhang, J.-G., Xu, W. & Henderson, W. A. *Lithium metal anodes and rechargeable lithium metal batteries*. 1 edn, 194 (Springer International Publishing, 2017).
3. Lin, D., Liu, Y. & Cui, Y. Reviving the lithium metal anode for high-energy batteries. *Nature Nanotech.* **12**, 194 (2017).
4. Wood, K. N., Noked, M. & Dasgupta, N. P. Lithium Metal Anodes: Toward an Improved Understanding of Coupled Morphological, Electrochemical, and Mechanical Behavior. *ACS Energy Lett.* **2**, 664-672 (2017).
5. Lin, D. *et al.* Layered reduced graphene oxide with nanoscale interlayer gaps as a stable host for lithium metal anodes. *Nature Nanotech.* **11**, 626 (2016).
6. Yang, C.-P., Yin, Y.-X., Zhang, S.-F., Li, N.-W. & Guo, Y.-G. Accommodating lithium into 3D current collectors with a submicron skeleton towards long-life lithium metal anodes. *Nature Commun.* **6**, 8058 (2015).
7. Bai, P. *et al.* Interactions between lithium growths and nanoporous ceramic separators. *Joule* **2**, 2434-2449 (2018).
8. Zhao, C.-Z. *et al.* An ion redistributor for dendrite-free lithium metal anodes. *Sci. Adv.* **4**, eaat3446 (2018).
9. Liu, Y. *et al.* Making Li-metal electrodes rechargeable by controlling the dendrite growth direction. *Nature Energy* **2**, 17083 (2017).
10. Zheng, J. *et al.* Electrolyte additive enabled fast charging and stable cycling lithium metal batteries. *Nature Energy* **2**, 17012 (2017).
11. Ren, X. *et al.* Guided lithium metal deposition and improved lithium Coulombic efficiency through synergistic effects of $LiAsF_6$ and cyclic carbonate additives. *ACS Energy Lett.* **3**, 14-19 (2018).
12. Zhang, X.-Q., Cheng, X.-B., Chen, X., Yan, C. & Zhang, Q. Fluoroethylene Carbonate Additives to Render Uniform Li Deposits in Lithium Metal Batteries. *Adv. Funct. Mater.* **27**, 1605989 (2017).
13. Qian, J. *et al.* Dendrite-free Li deposition using trace-amounts of water as an electrolyte additive. *Nano Energy* **15**, 135-144 (2015).
14. Kim, M. S. *et al.* Langmuir–Blodgett artificial solid-electrolyte interphases for practical lithium metal batteries. *Nature Energy* **3**, 889-898 (2018).
15. Li, N.-W., Yin, Y.-X., Yang, C.-P. & Guo, Y.-G. An artificial solid electrolyte interphase layer for stable lithium metal anodes. *Adv. Mater.* **28**, 1853-1858 (2016).
16. Tu, Z. *et al.* Fast ion transport at solid–solid interfaces in hybrid battery anodes. *Nature Energy* **3**, 310-316 (2018).
17. Liang, X. *et al.* A facile surface chemistry route to a stabilized lithium metal anode. *Nature Energy* **2**, 17119 (2017).
18. Suo, L. *et al.* Fluorine-donating electrolytes enable highly reversible 5-V-class Li metal batteries. *Proceedings of the National Academy of Sciences* **115**, 1156-1161 (2018).

19. Fan, X. *et al.* Highly Fluorinated Interphases Enable High-Voltage Li-Metal Batteries. *Chem* **4**, 174-185 (2018).
20. Qian, J. *et al.* High rate and stable cycling of lithium metal anode. *Nature Commun.* **6**, 6362 (2015).
21. Han, X. *et al.* Negating interfacial impedance in garnet-based solid-state Li metal batteries. *Nature Mater.* **16**, 572 (2016).
22. Duan, H. *et al.* Dendrite-free Li-metal battery enabled by a thin asymmetric solid electrolyte with engineered layers. *J. Am. Chem. Soc.* **140**, 82-85 (2018).
23. Li, J., Ma, C., Chi, M., Liang, C. & Dudney, N. J. Solid Electrolyte: the Key for High-Voltage Lithium Batteries. *Adv. Energy Mater.* **5**, 1401408 (2015).
24. Bai, P., Li, J., Brushett, F. R. & Bazant, M. Z. Transition of lithium growth mechanisms in liquid electrolytes. *Energy Environ. Sci.* **9**, 3221-3229 (2016).
25. Kushima, A. *et al.* Liquid cell transmission electron microscopy observation of lithium metal growth and dissolution: Root growth, dead lithium and lithium flotsams. *Nano Energy* **32**, 271-279 (2017).
26. Li, Y. *et al.* Atomic structure of sensitive battery materials and interfaces revealed by cryo–electron microscopy. *Science* **358**, 506-510 (2017).
27. Li, Y. *et al.* Correlating structure and function of battery interphases at atomic resolution using cryoelectron microscopy. *Joule* **2**, 2167-2177 (2018).
28. Lu, Y., Tu, Z. & Archer, L. A. Stable lithium electrodeposition in liquid and nanoporous solid electrolytes. *Nature Mater.* **13**, 961 (2014).
29. Markevich, E., Salitra, G., Chesneau, F., Schmidt, M. & Aurbach, D. Very Stable Lithium Metal Stripping–Plating at a High Rate and High Areal Capacity in Fluoroethylene Carbonate-Based Organic Electrolyte Solution. *ACS Energy Lett.* **2**, 1321-1326 (2017).
30. Lu, Y., Tu, Z., Shu, J. & Archer, L. A. Stable lithium electrodeposition in salt-reinforced electrolytes. *J. Power Sources* **279**, 413-418 (2015).
31. Wu, F. *et al.* Lithium iodide as a promising electrolyte additive for lithium–sulfur batteries: mechanisms of performance enhancement. *Adv. Mater.* **27**, 101-108 (2015).
32. Yan, C. *et al.* Lithium nitrate solvation chemistry in carbonate electrolyte sustains high-voltage lithium metal batteries. *Angew. Chem. Int. Ed.* **57**, 14055-14059 (2018).
33. Li, W. *et al.* The synergetic effect of lithium polysulfide and lithium nitrate to prevent lithium dendrite growth. *Nature Commun.* **6**, 7436 (2015).
34. Ma, L., Kim, M. S. & Archer, L. A. Stable artificial solid electrolyte interphases for lithium batteries. *Chem. Mater.* **29**, 4181-4189 (2017).
35. Ishikawa, M., Morita, M. & Matsuda, Y. In situ scanning vibrating electrode technique for lithium metal anodes. *J. Power Sources* **68**, 501-505 (1997).
36. Kazyak, E., Wood, K. N. & Dasgupta, N. P. Improved Cycle Life and Stability of Lithium Metal Anodes through Ultrathin Atomic Layer Deposition Surface Treatments. *Chem. Mater.* **27**, 6457-6462 (2015).
37. Liu, Y. *et al.* An artificial solid electrolyte interphase with high Li-Ion conductivity, mechanical strength, and flexibility for stable lithium metal anodes. *Adv. Mater.* **29**, 1605531 (2017).
38. Weinstein, L., Yourey, W., Gural, J. & Amatucci, G. G. Electrochemical impedance spectroscopy of electrochemically self-assembled lithium–iodine batteries. *J. Electrochem. Soc.* **155**, A590-A598 (2008).
39. Behl, W. K. & Chin, D. T. Electrochemical overcharge protection of rechargeable lithium batteries: I. Kinetics of iodide/tri-Iodide/iodine redox reactions on platinum in solutions. *J. Electrochem. Soc.* **135**, 16-21 (1988).

40. Behl, W. K. & Chin, D. T. Electrochemical overcharge protection of rechargeable lithium batteries: II. Effect of lithium iodide-iodine additives on the behavior of lithium electrode in solutions. *J. Electrochem. Soc.* **135**, 21-25 (1988).
41. Behl, W. K. Anodic oxidation of lithium bromide in tetrahydrofuran solutions. *J. Electrochem. Soc.* **136**, 2305-2310 (1989).
42. Xianming, W., Yasukawa, E. & Kasuya, S. Electrochemical properties of tetrahydropyran-based ternary electrolytes for 4 V Lithium metal rechargeable batteries. *Electrochimica Acta* **46**, 813-819 (2001).
43. Yamaki, J.-I. The development of lithium rechargeable batteries. *J. Power Sources* **20**, 3-7 (1987).
44. Zhang, S. S. Role of LiNO3 in rechargeable lithium/sulfur battery. *Electrochimica Acta* **70**, 344-348 (2012).
45. Ding, F. *et al.* Effects of Carbonate Solvents and Lithium Salts on Morphology and Coulombic Efficiency of Lithium Electrode. *J. Electrochem. Soc.* **160**, A1894-A1901 (2013).
46. Han, F. *et al.* High electronic conductivity as the origin of lithium dendrite formation within solid electrolytes. *Nature Energy* (2019).
47. Blöchl, P. E. Projector augmented-wave method. *Phys. Rev. B* **50**, 17953-17979 (1994).
48. Kresse, G. & Joubert, D. From ultrasoft pseudopotentials to the projector augmented-wave method. *Phys. Rev. B* **59**, 1758-1775 (1999).
49. Mortensen, J. J., Hansen, L. B. & Jacobsen, K. W. Real-space grid implementation of the projector augmented wave method. *Phys. Rev. B* **71**, 035109 (2005).
50. Enkovaara, J. *et al.* Electronic structure calculations with GPAW: a real-space implementation of the projector augmented-wave method. *J. Phys. Cond. Mater.* **22**, 253202 (2010).
51. Perdew, J. P., Burke, K. & Ernzerhof, M. Generalized gradient approximation made simple. *Phys. Rev. Lett.* **77**, 3865-3868 (1996).
52. Monkhorst, H. J. & Pack, J. D. Special points for Brillouin-zone integrations. *Phys. Rev. B* **13**, 5188-5192 (1976).
53. Tang, W., Sanville, E. & Henkelman, G. A grid-based Bader analysis algorithm without lattice bias. *J. Phys. Cond. Mater.* **21**, 084204 (2009).

# Supplementary Information for

**Design Principles for Self-forming Interfaces Enabling Stable Lithium Metal Anodes**

Yingying Zhu,[∟, ⌈] Vikram Pande,[⊥, ⌈] Linsen Li,[∟, †, *] Sam Pan,[†] Bohua Wen,[†] David Wang,[†] Venkatasubramanian Viswanathan,[⊥, *] Yet-Ming Chiang[†, *]

[∟] Department of Chemical Engineering, Shanghai Jiao Tong University, Shanghai, China 200240

[†] Department of Materials Science & Engineering, Massachusetts Institute of Technology, Cambridge, MA 02139, USA

[⊥] Department of Mechanical Engineering, Carnegie Mellon University, Pittsburgh PA 15213，USA

[⌈] These authors contribute equally to the work.

* Corresponding authors

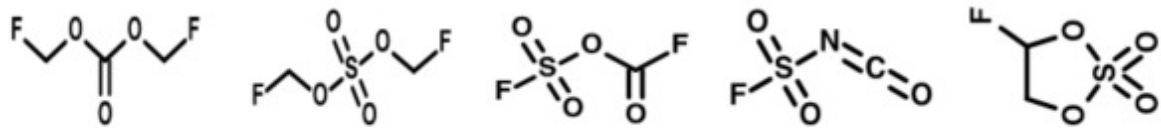

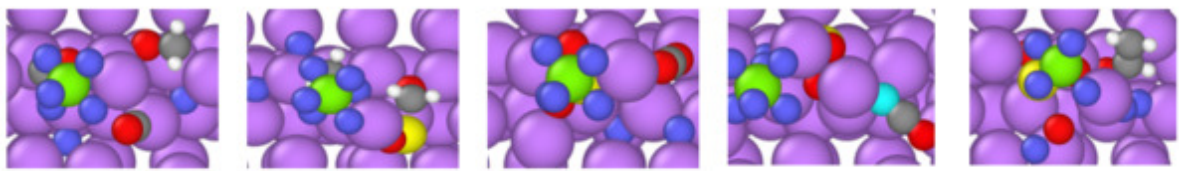

**Molecular Structures**

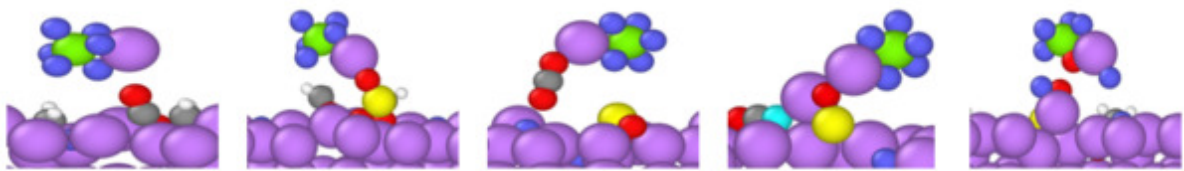

**Top View**

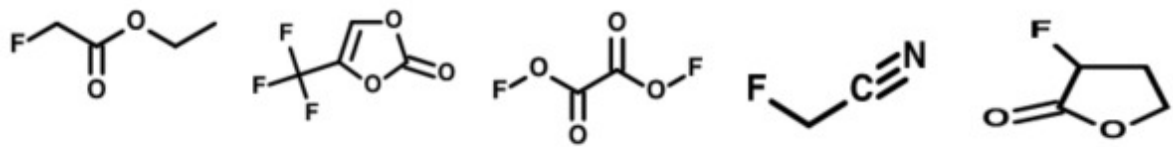

**Side View**

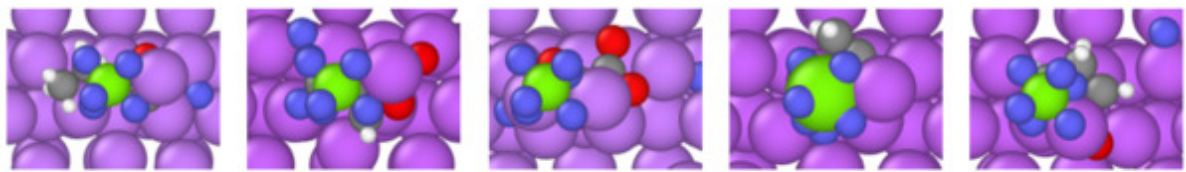

**Molecular Structures**

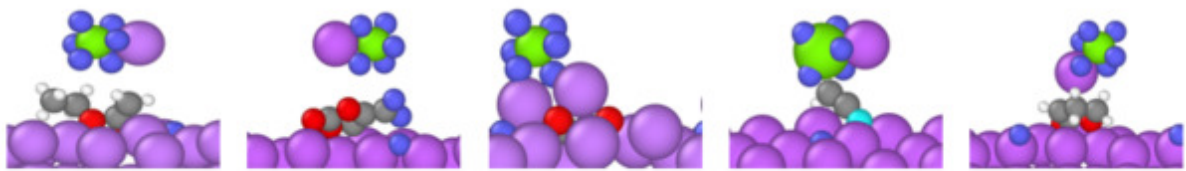

**Top View**

**Side View**

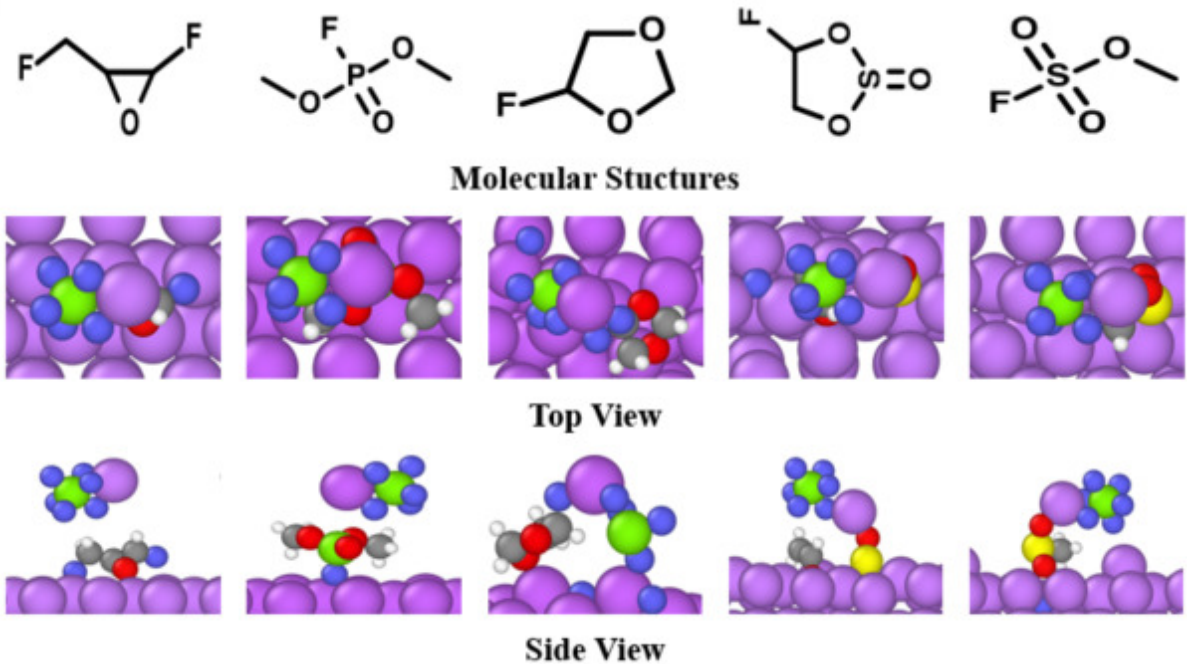

**Supplementary Figure 1 | Decomposition of fluorinated compounds at Li surface predicted by DFT calculations.** We show here the decomposition of 15 compounds. Most fluorinated compounds do lead to the formation of LiF, but there are some compounds which do not release the fluorine such as fluorinated epoxides, ethers and compounds with $CF_3$ and $CF_2$ groups. We also see that some compounds such as fluorinated DTD, fluorinated dioxane lead to the formation of additional LiF by decomposing the $LiPF_6$ salt. Lastly all sulfate and sulfur groups decompose readily to form $SO_2^{2-}$ anion which we believe would ultimately lead to the formation of $Li_2SO_3$ and $ROSO_2Li$.

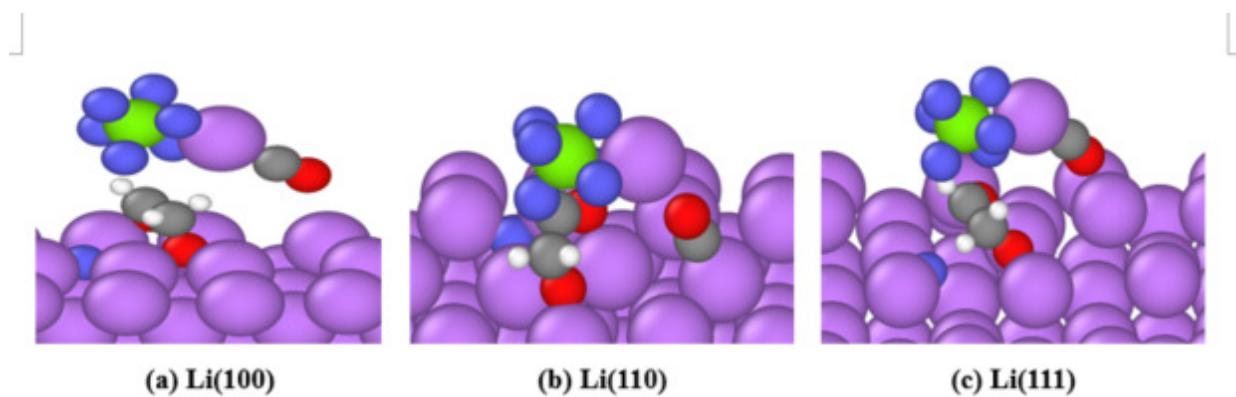

**Supplementary Figure 2 | Decomposition of FEC in the presence of LiPF$_6$ salt on Li (100), (110) and (111) surfaces**. The pink atoms represent Li, green represent P and purple represent F. The decomposition products are identical in all cases. We suggest that the surface energy effects do not significantly affect the decomposition pathway for solvent decomposition on Li. We also see identical results for other solvents considered in this study. We hypothesize that the reason behind this is surface energy difference between different surface ~0.2–0.5 eV while the energy difference is an order of magnitude higher for the decomposition reaction ~2–5 eV. This shows that even during SEI healing, the exposed Li cracks will react to give similar reaction products assuming there are sufficient free Li atoms to complete the decomposition. In some cases, it is possible that due to passivation, the complete decomposition does not happen.

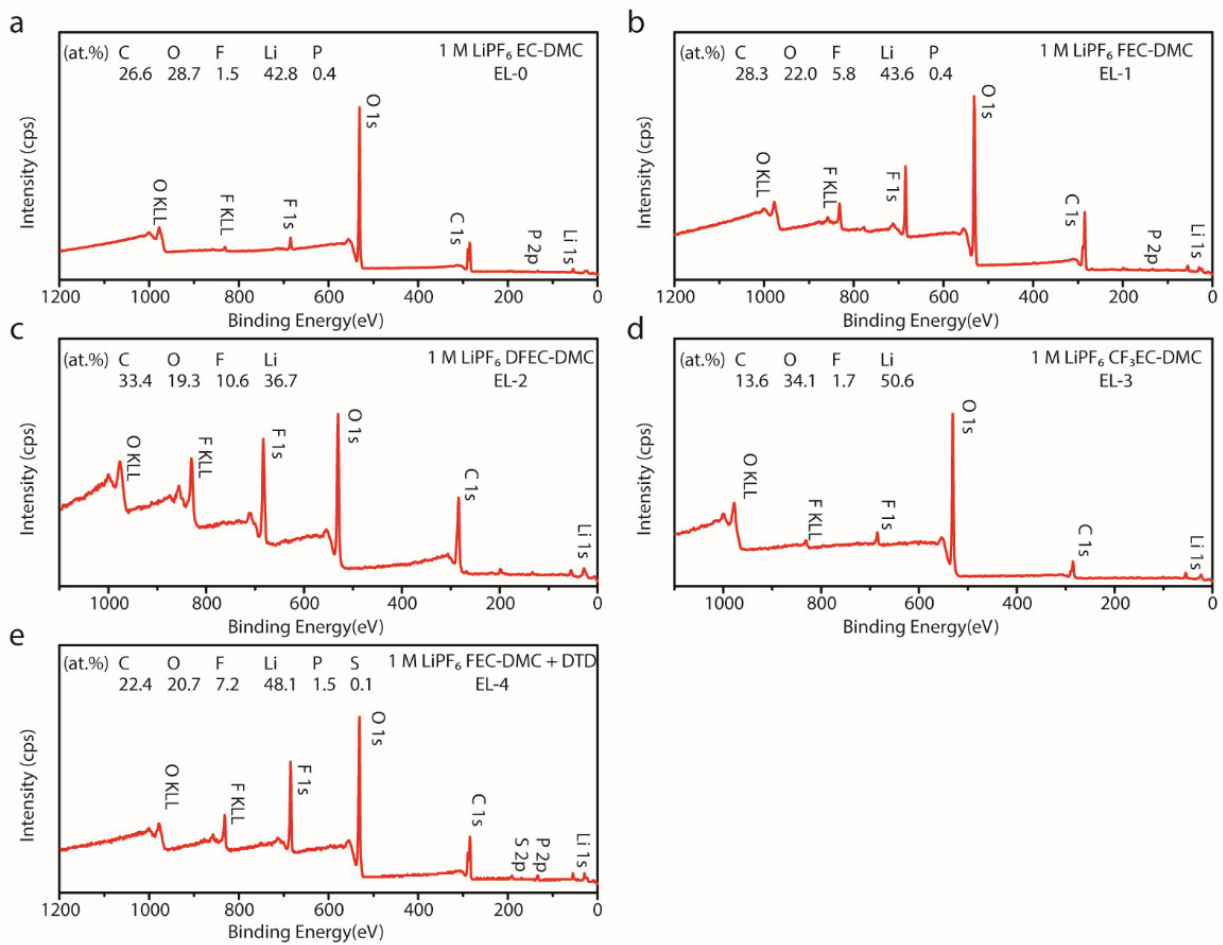

**Supplementary Figure 3 | Wide-scan XPS spectra of five different electrolytes.** Wide-scan XPS spectra of the Li metal SEI formed in **a.** 1 M LiPF$_6$ EC-DMC (EL-0), **b.** 1 M LiPF$_6$ FEC-DMC (EL-1), **c.** 1 M LiPF$_6$ DFEC-DMC (EL-2), **d.** 1 M LiPF$_6$ CF$_3$EC-DMC (EL-3), and **e.** 1 M LiPF$_6$ FEC-DMC + DTD (EL-4). The fluorine content in the SEI increases in the order of EL-0, EL-4, EL-1, EL-4, and EL-2.

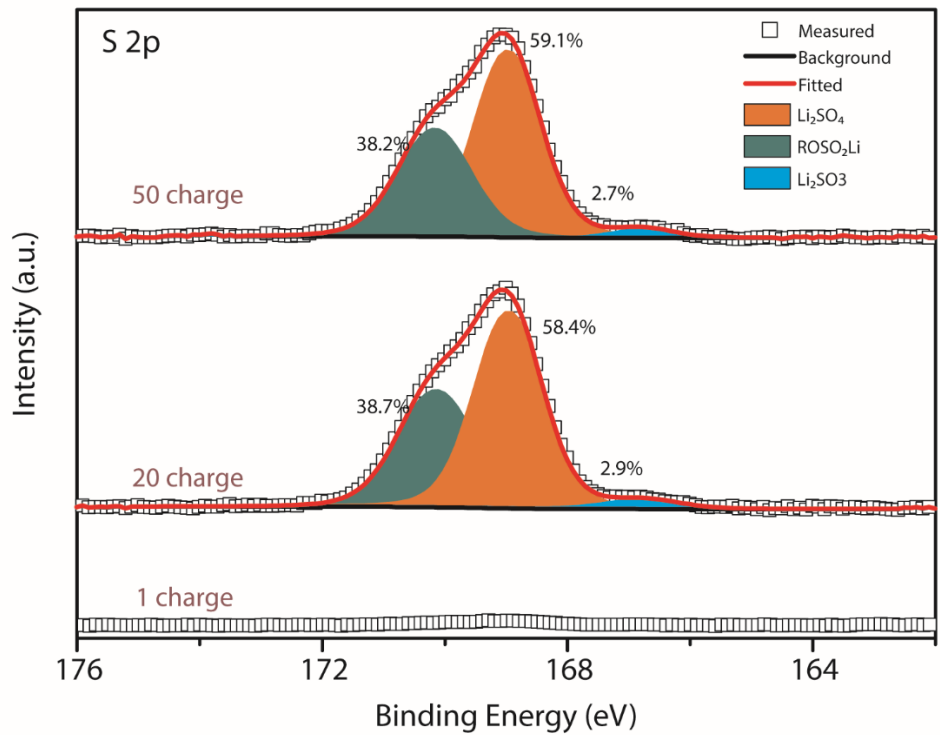

**Supplementary Figure 4 | Narrow-scan XPS spectra of S 1s and peak analysis.** Narrow-scan XPS spectra of S 1s collected from the SEI formed on the Li electrode surface at the 1st charge, 20th charge, and 50th charge cycle. There is little S (0.1 at.%) in the SEI formed at the 1st charge. S content increases to 1.7 at.% and 2.2 at.% at the 20th and 50th charge, respectively. Based on peak analysis, there are three different S species, namely $Li_2SO_4$, $ROSO_2Li$, and a small amount of $Li_2SO_3$.

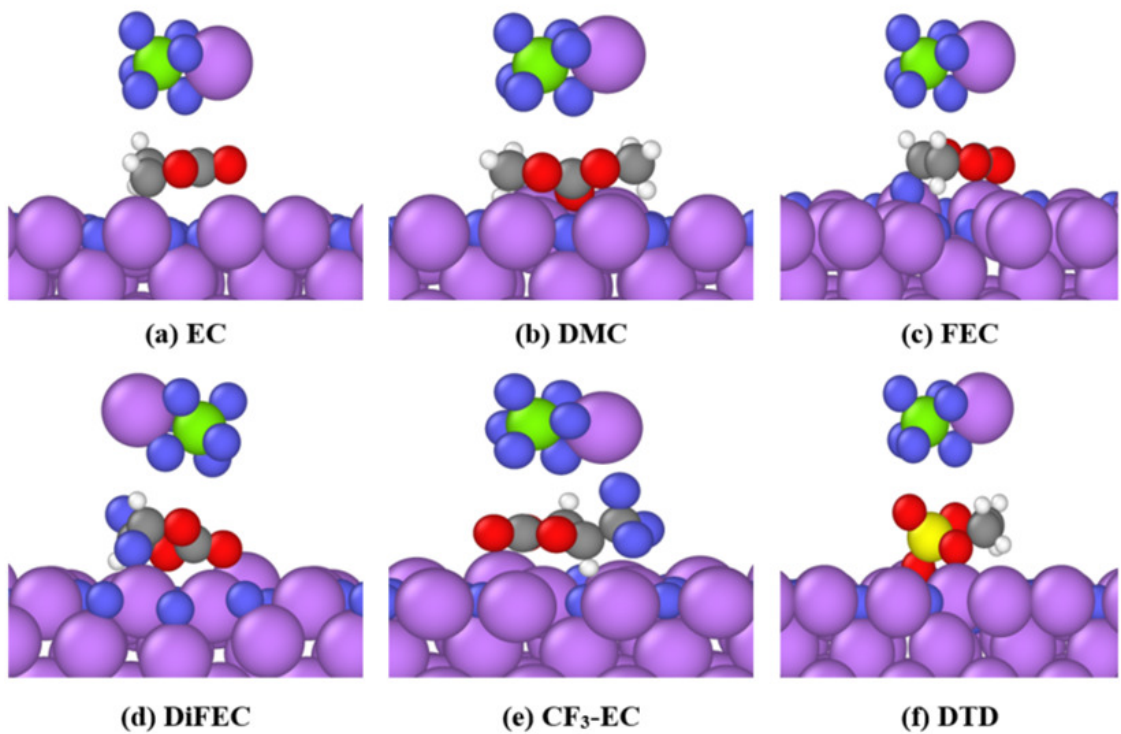

**Supplementary Figure 5 | Decomposition of solvent molecules on Li(100) surface with one mono layer of LiF.** The pink atoms represent Li, red represent O, grey represent C, green represent P, purple represent F, yellow represent S and white represent H. There is no chemical decomposition of the solvent in any of the cases. This is also validated from the electrons transferred from the Li slab which is less than 0.5 electrons as shown in Supplementary Table 1. Even in the case of DTD, the co-decomposition of DTD and $LiPF_6$ is stopped due to the unavailability of Li. This clearly shows that a monolayer of LiF is sufficient to chemically passivate a surface from further solvent decomposition (chemically). This proves that in general increased fraction of LiF in the SEI will lead to a more compact and dense SEI.

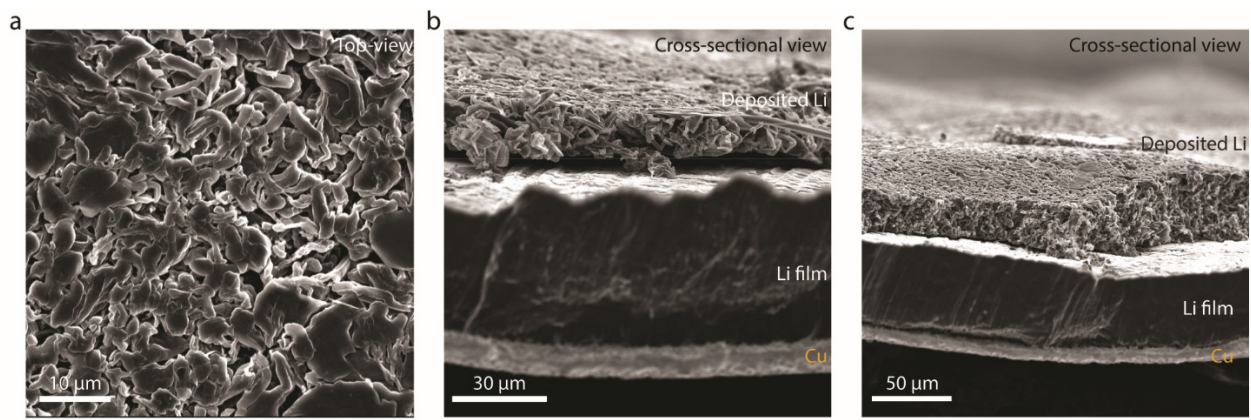

**Supplementary Figure 6 | SEM characterization of the deposited Li film on the Li/Cu substrates from Li-Li cells. a**, Top-view SEM images of the deposited Li film on the Li/Cu substrates (50 μm-thick Li, 15 μm-thick Cu) in 1 M $LiPF_6$ EC-DMC (EL-0). The Li film was deposited in a Li-Li (thick Li foil vs 50 μm-thick Li on Cu) cell by charging at 0.42 mA $cm^{-2}$ for 10 h. 4.2 mAh $cm^{-2}$ of Li was deposited. More whisker-like Li particles were observed when a thick Li foil was used as the counter electrode instead of the roller-pressed $LiCoO_2$ electrode shown in Figure 3 of the main text. **b** and **c** are the corresponding cross-sectional SEM images, showing three-layer structure consisting of the deposited Li, the original 50 μm-thick Li, and underlying Cu substrate. The deposited Li film consisted of whisker-like Li particles and had a wave-like surface, making it difficult to measure the thickness of the deposited Li film. The surface of the Li films deposited using $LiCoO_2$-Li cells were more flat, as shown in Figure 3 of the main text.

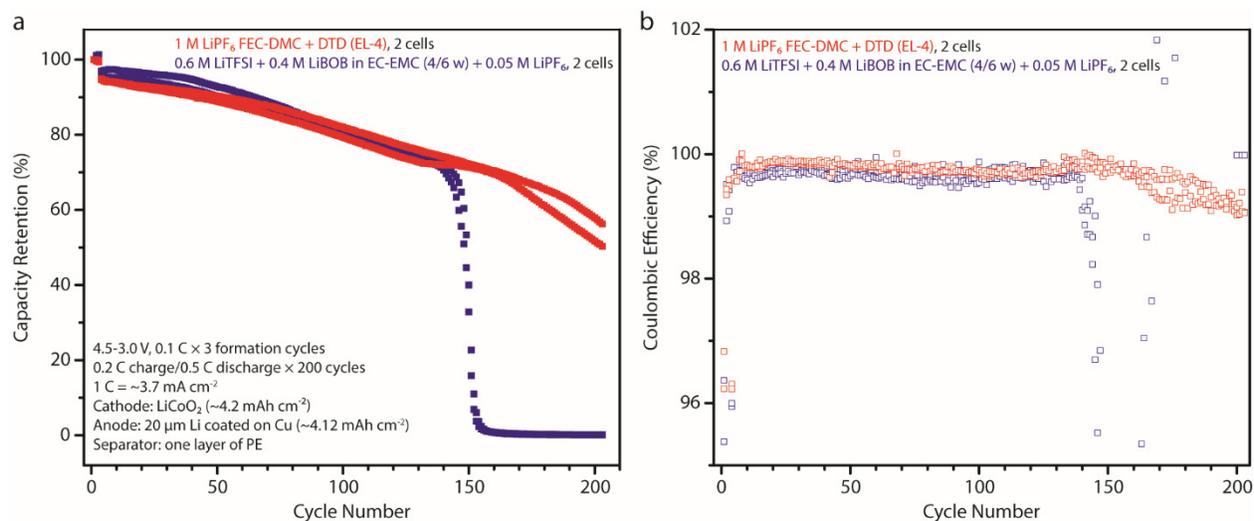

**Supplementary Figure 7 | Electrochemical tests of LiCoO$_2$-Li full batteries. a,** The cycling performance of the LiCoO$_2$-Li full cells using the EL-4 electrolyte (1 M LiPF$_6$ FEC-DMC + 3wt% DTD) shown in comparison with those using the dual salt electrolyte (0.6 M LiTFSI + 0.4 M LiBOB in EC-EMC 4/6 by weight + 0.05 M LiPF$_6$) previously reported in Zheng, J. *et al. Nature Energy* **2**, 17012 (2017). The cells were tested under the same conditions. Data from two cells was shown in each case. The EL-4 electrolyte provides better cycling performance than the dual-salt electrolyte. **b,** Coulombic efficiency of the four cells. The EL-4 electrolyte provides higher CE than the dual-salt electrolyte.

**Supplementary Tables**

**Supplementary Table 1 | Theoretical descriptors for solvent decomposition on a mono layer of LiF formed on Li(100)**

| Solvent | Electrons Transferred on LiF |
|---|---|
| DMC | 0.3 |
| EC | 0.2 |
| FEC | 0.2 |
| DiFEC | 0.2 |
| $CF_3$-EC | 0.1 |
| DTD | 0.3 |

**Supplementary Table 2 | Average Coulombic efficiency of the thin Li anodes in different electrolytes**

| Lithium Salt (1 M) | Solvents & Additives | $CE_{avg}$ |
|---|---|---|
| $LiPF_6$ | PC | 73.6% |
| LiTFSI | DOL-DME (1:1 v), 0.1 M LiI | 83.0% |
| $LiPF_6$ | EC-EMC (1:1 v) | 90.1% |
| LiTFSI | EC-THP (1:1 v) | 95.5% |
| $LiAsF_6$ | EC-DMC (1:1 v) | 95.9% |
| $LiAsF_6$ | EC-2MeTHF (1:1 v) | 96.4% |
| LiTFSI | DOL-DME (1:1 v), 1 wt% $LiNO_3$ | 96.7% |

**Test conditions**: 0.6 mA $cm^{-2}$ current density, 3.0 mAh $cm^{-2}$ cycling areal capacity. PC = propylene carbonate, DOL = 1, 3-dioxolane, DME = 1,2-Dimethoxyethane, EMC = ethyl methyl carbonate, THP = tetrahydropyran, 2-MeTHF = 2-methyl tetrahydrofuran

Supplementary Table 3 is a separate Microsoft Excel file.

**Supplementary Text**

**DFT Calculation Details**

Self-Consistent DFT calculations were performed using the real space projector-augmented wave method[1,2] implemented in the GPAW code[3,4] and employing the PBE exchange correlation functional[5]. We chose the Li (100), (110) and (111) surfaces for the DFT calculations. The Li surface comprised of four layers with the bottom two layers constrained at the bulk lattice constants. Each layer consisted of 3x3 Li unit cell. The solvent molecule was placed on top of Li surface at a distance of 2 Å. We explored different conformers of the solvent molecule, chosen based on placing electronegative atoms such as F and O close to the surface. $Li^+$ and $PF_6^-$ ions were placed at a distance of 2 Å on top of the solvent molecule. $LiPF_6$ salt was placed for decomposition studies as it is known that salt ions affect the stability of solvent molecules by renormalizing the molecular energy levels of the solvent.[6,7] In addition, the salt ions may themselves participate in the reaction. It is worth highlighting that there are numerous possible configurations of the salt ions and solvent, we believe that given the consistency between the structures, trends in reactivity are well captured with this approach. The internal coordinates of these structures were allowed to relax to determine the decomposition products. Periodic boundary conditions were used for x and y directions and a vacuum of 10 Å was used in the z direction perpendicular to the surface on both sides of the slab. A real-space grid spacing of 0.16 Å was used and the Brillouin zone was sampled using the Monkhorst Pack scheme[8] with a k-point grid of (6×6×1). The calculations were converged to < 5meV accuracy with respect to k-points and grid spacing. A Fermi-Dirac smearing of 0.05 eV All simulations were converged to a force < 0.05 eV Å$^{-1}$. Bader analysis[9] was used to determine the amount of charge transferred from the lithium to the solvent during the decomposition and also volumes of the various decomposed species.

**Calculation of Bader Charges and Bader Volumes of different species**

To calculate the Bader charges and volumes, initially the electron density as a function of spatial coordinates was stored in a ".cube" file. The Bader analysis was performed on the ".cube" file. For charges, the Bader analysis was done by setting the vacuum charge density to zero. This was done to ensure that all charges are assigned to the appropriate molecular species. For calculating Bader Volumes of the atoms, the Bader analysis was done by setting the vacuum charge density to 0.0001 $e$ Å$^{-3}$. This was the error of the electron density in the DFT calculations performed. Thus, a cutoff lower than this value would not be consistent. A larger value for the vacuum charge density cutoff leads to incorrect assignment of the electrons to different atoms. We checked for some different values in the appropriate range and found that the Bader Volumes calculated scale with those chosen number, but the trends in the volumes of different species remain the same as shown in the table shown below. This means that the specie with the largest Bader volume is invariant implying that the descriptor used will provide the correct trend for the classification problem. The Bader charge transferred to the solvent while decomposition was calculated as negative of the charge on the Li(100) slab because the overall system is charge neutral. The decomposed species were identified by considering bond distances between two atoms. Two atoms were considered chemically bonded if the distance between them was less than 1.75 Å. Thus, all the decomposed species were identified and their charge and volumes calculated by summing over the charges and volumes of the individual atoms.

| Solvent | Bader Volume ($Å^3$) for Cutoff Density | | |
|---|---|---|---|
| | 0.0001 (e/$Å^3$) | 0.0005 (e/$Å^3$) | 0.001 (e/$Å^3$) |
| DMC | 145 | 120 | 107 |
| EC | 132 | 110 | 99 |
| FEC | 89 | 75 | 67 |
| DiFEC | 125 | 103 | 92 |
| $CF_3$-EC | 169 | 137 | 122 |
| DTD | 97 | 81 | 74 |

**Passivation of Li(100) surface covered with LiF**

In order to explore the extent of passivation by LiF, we performed calculations by placing a monolayer of LiF on a 6 layer Li(100) slab. The structure was generated by placing F atoms on top of the Li surface. After relaxation, this spontaneously led to the formation of a LiF monolayer and 5 layers of Li(100). For simulating solvent decomposition on this structure, the bottom two layer were constrained to the bulk lattice constant of Li and the other layers and solvent molecule along with $LiPF_6$ was allowed to relax. A similar Bader charge analysis was done to determine the charge transferred to the solvent. This shows us whether there is any Li oxidation and hence any possible reaction with the solvent. If the charge transferred is less than 0.5 $e^-$ then, this would indicate that the surface is passivated and no further reaction with the solvent is likely.

**Calculation of Energy Density in Figure 1**

| | LCO mass (mg) | Separator mass (mg) | Li (mg) | Cu (mg) | Al (mg) | Electrolyte (mg) | Total Mass (mg) | Energy (Wh) | Energy Density (Wh/kg) | Total Volume (cm^3) | Volumetric Energy Density (Wh/L) |
|---|---|---|---|---|---|---|---|---|---|---|---|
| LCO-Li (20um Li) | 29.13 | 3.80 | 1.35 | 8.51 | 2.57 | 4.18 | 49.55 | 0.02156 | 370 | 0.0178 | 1029 |
| LCO-Li (50 um Li) | 29.13 | 3.80 | 3.38 | 8.51 | 2.57 | 4.18 | 51.58 | 0.02156 | 355 | 0.0216 | 848 |
| LCO-Li (100 um Li) | 29.13 | 3.80 | 6.76 | 8.51 | 2.57 | 4.18 | 54.96 | 0.02156 | 333 | 0.0280 | 656 |
| LCO-Li (250 um Li) | 29.13 | 3.80 | 16.91 | 8.51 | 2.57 | 4.18 | 65.10 | 0.02156 | 281 | 0.0470 | 390 |
| LCO-Li (450 um Li) | 29.13 | 3.80 | 30.44 | 8.51 | 2.57 | 4.18 | 78.63 | 0.02156 | 233 | 0.0723 | 253 |
| LCO-Li (750 um Li) | 29.13 | 3.80 | 50.73 | 8.51 | 2.57 | 4.18 | 98.93 | 0.02156 | 185 | 0.1103 | 166 |
| LCO-Cu (anode free) | 29.13 | 3.80 | 0.00 | 8.51 | 2.57 | 4.18 | 48.19 | 0.02156 | 380 | 0.0153 | 1199 |

| | | Capacity [mAh/g] | Density [g/cm3] | porosity | thickness [um] |
|---|---|---|---|---|---|
| Cathode | LCO | 185 | 5 | 0.25 | 60 |
| Anode | Lithium | 3860 | | 0 | |
| Electrolyte | EC-EMC | | 1.2 | | |
| Current collector | Cu | | 8.96 | | 15 |
| | Al | | 2.7 | | 15 |
| Separator | | | 1.2 | 0.5 | 25 |

Packaging factor = 0.85

Electrode area = 1.267 cm$^2$

**Supplementary References**

[1] Blöchl, P. E., Projector augmented-wave method. ***Physical review B***, 50(24), 17953 (1994).

[2] Kresse, G., & Joubert, D., From ultrasoft pseudopotentials to the projector augmented-wave method. ***Physical Review B***, 59(3), 1758 (1999).

[3] Mortensen, J. J., Hansen, L. B., & Jacobsen, K. W., Real-space grid implementation of the projector augmented wave method. ***Physical Review B***, 71(3), 035109 (2005).

[4] Enkovaara, J. E., Rostgaard, C., Mortensen, J. J., Chen, J., Dułak, M., Ferrighi, L. & Kristoffersen, H. H., Electronic structure calculations with GPAW: a real-space implementation of the projector augmented-wave method. ***Journal of Physics: Condensed Matter***, 22(25), 253202 (2010).

[5] Perdew, J. P., Burke, K., & Ernzerhof, M., Generalized gradient approximation made simple.***Phys. Rev. Lett.***, 77(18), 3865 (1996).

[6] Kumar, N., & Siegel, D. J. (2016). Interface-induced renormalization of electrolyte energy levels in magnesium batteries. ***J. Phys. Chem. Lett.***, 7(5), 874-881.

[7] Khetan, A., Luntz, A., & Viswanathan, V. (2015). Trade-offs in capacity and rechargeability in nonaqueous Li–$O_2$ batteries: Solution-driven growth versus nucleophilic stability. ***J. Phys. Chem. Lett.***,, 6(7), 1254-1259.

[8] Monkhorst, H. J., & Pack, J. D. Special points for Brillouin-zone integrations. ***Physical review B***, 13(12), 5188 (1976).

[9] Tang, W., Sanville, E., & Henkelman, G., A grid-based Bader analysis algorithm without lattice bias. ***Journal of Physics: Condensed Matter***, 21(8), 084204 (2009).